\documentclass[11pt,twoside]{article}
\usepackage{asp2010}

\resetcounters

\begin{document}

\title{Compression of Visibility Data for Murchison Widefield Array.}
\author{Vyacheslav V. Kitaeff$^1$
\affil{$^1$International Centre for Radio Astronomy Research, The University of Western Australia, M468, 35 Stirling Hwy, Crawley 6009, WA, Australia, +61-8-64887744, slava.kitaeff@icrar.org}}

\begin{abstract}
The Murchison Widefield Array (MWA) is a new low frequency radio telescope 
operating on the Square Kilometre Array site in Western Australia. MWA
is generating tens of terabytes of data daily. The size of the required data storage
has become a significant operational limitation and cost. We present a simple binary 
compression technique and a system for the floating point visibility data 
developed MWA. We present the statistics of the impact of such 
compression on the data with the typical compression ratio up to 1:3.1.
\end{abstract}

\section{Introduction}

The MWA is one of the Square Kilometre Array (SKA) 
precursor radio telescopes located in the desert in Western Australian. MWA is an interferometer
type radio telescope with the longest baseline $2.9$ km. It is optimised for the observations of 
very wide fields of sky between 200 and 2500 square degrees, and the frequencies 
between 80 and 300 MHz, with a processed bandwidth $30.72 MHz$ for both linear polarisations
  \citep{TINGAY}. The MWA works without any moving parts, 
the primary beam is formed by electronically adjusted phases of the 16 bow-tie dipole antennas
that form a small phased array known as a ÒtileÓ. The complete array of 128 tiles provides 
about 2000 $m^2$ of collecting area at 150 MHz.

\begin{figure}
  \centering
  \includegraphics[width=120mm]{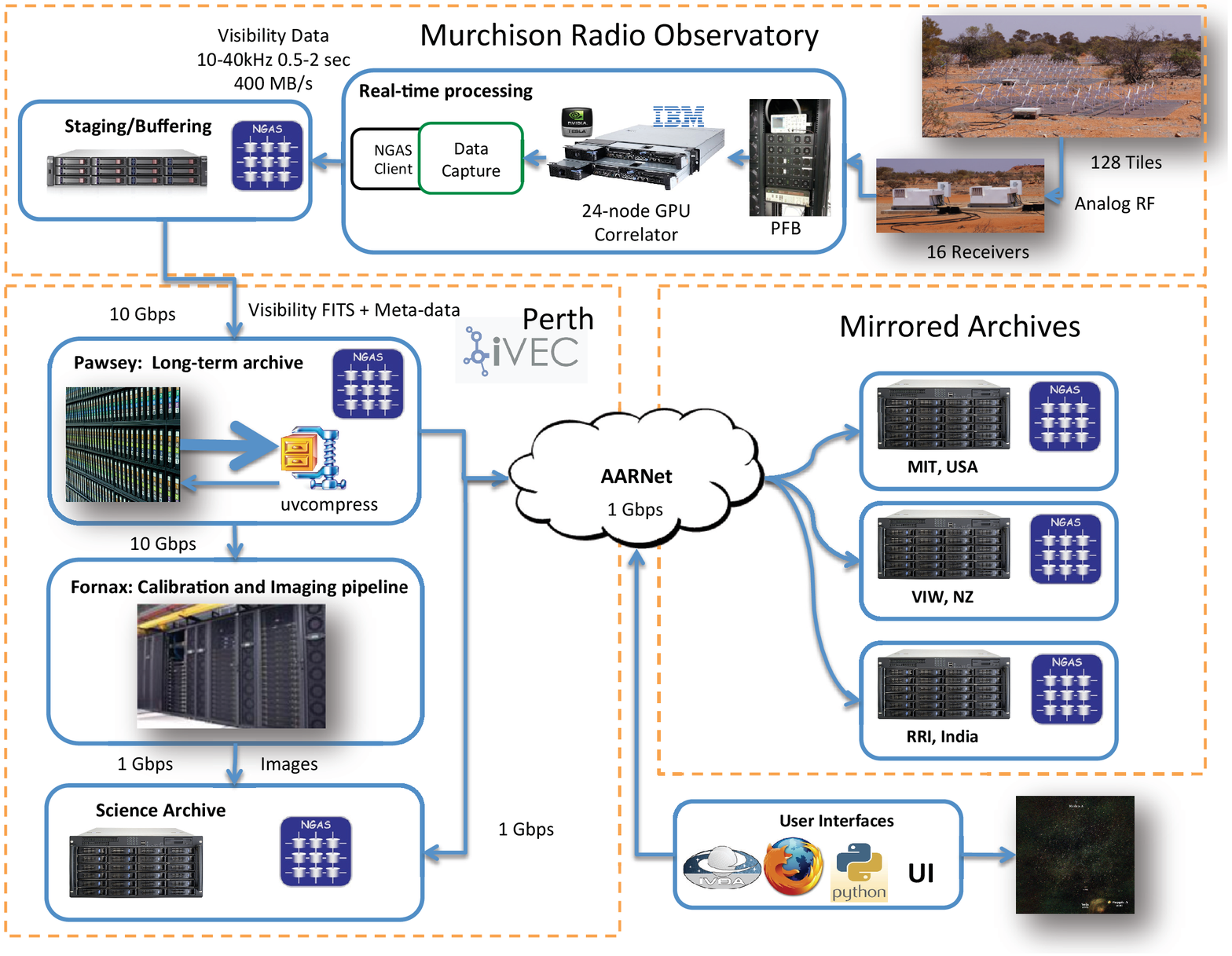}
  \caption{MWA data management system.}
  \label{fig:mwa}
\end{figure}

Figure~\ref{fig:mwa} shows the generic data flow in MWA digital backend system. The MWA implements 
a hybrid, distributed FX correlator outputting complex visibilities at 10--40 $KHz$
frequency resolution integrated to 0.5--2 sec as required.

Each visibility is a complex number representing the amplitude and phase of
a signal at a particular point in time within a frequency channel. For a given
channel at each time step, the correlator carries out $2N\times(N+1)$ pair-wise 
cross-correlation and auto-correlation \citep{2009PASP..121..857W}, where $N$ is
the number of tiles. Since a visibility is a complex number, it is stored as 2 single precision 
(32 bit) IEEE 754 floating point data type values. The typical data rate of visibility data, for example, 
for $30.72 MHz$ bandwidth, $10 kHz$ spectral resolution and 2 second integration time is 387 MB/s. 
The Data Capture software component captures the data, forms memory-resident 
$24\times$ FITS files for each correlator dump, and passes the files to NGAS-Client \citep{2013ExA....36..679W}.
The data is then transferred via a dedicated optical fibre to Perth, and then further redistributed to other 
archives.

\section{Floating Point Errors and Compression}

During the F-stage of  correlation the integer input from the telescope's hardware is converted 
to floating point output due to the division operations. Such conversion is not precise due to the limitations
of IEEE 754 format. For example, a division by a prime number 127 can not be stored in IEEE 754 format
with an infinite precision. Such errors have a systematic nature, and propagate and accumulate further
down in the processing pipeline.

\begin{figure}
  \centering
  \includegraphics[width=100mm]{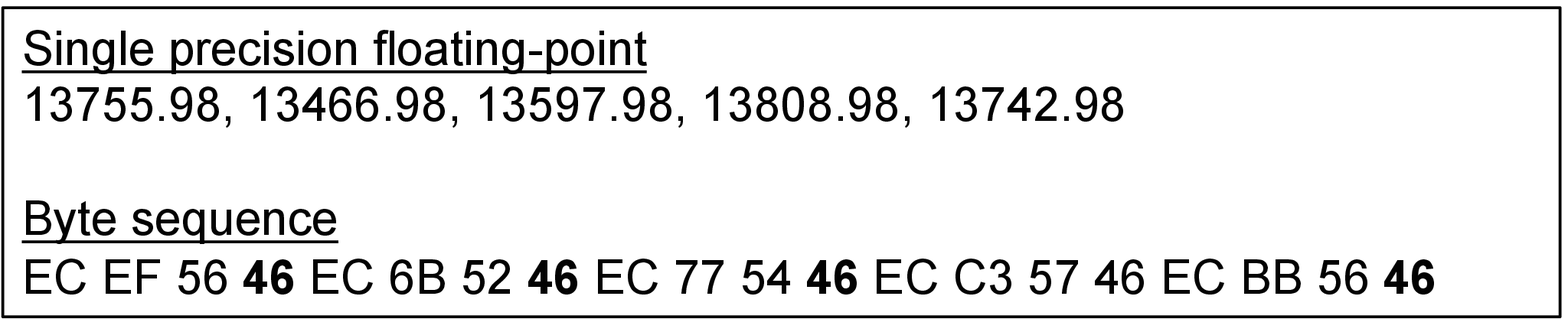}
  \caption{A fragment of floating point output of MWA correlator. The second line represents 
  the byte sequence of how the values are stored in a computer. Only the bytes designated in bold can be
  efficiently compressed with a binary algorithm.}
  \label{fig:original}
\end{figure}

Furthermore, IEEE 754 floating point format is poorly compressible. 
Figure~\ref{fig:original} shows a fragment of floating point output of MWA correlator. Only the bytes in bold
can be compressed with a binary compression algorithm, which typically gives not more than 25\% compression.
At the same time, usage of IEEE 754 floating point format creates a significant redundancy in the form of
storing the data whose origin is integer.

\begin{figure}
  \centering
  \includegraphics[width=120mm]{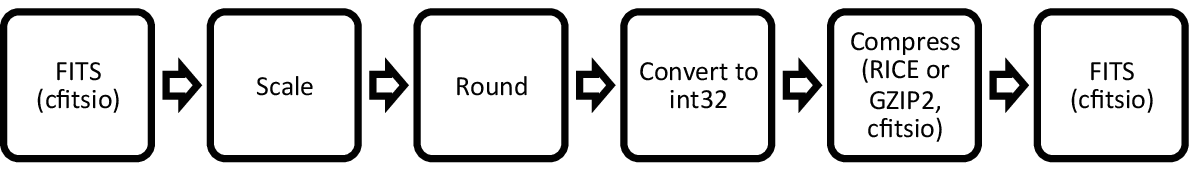}
  \caption{uvcompress algorithm.}
  \label{fig:algorithm}
\end{figure}

Recognising this, we have developed a simple but efficient compression algorithm specifically for MWA visibility data
dubbed \emph{uvcompressed}. Figure~\ref{fig:algorithm} outlines the steps in the algorithm.

First, all the values scaled by an optimal factor that is typically ranges from 1 to 1000. The optimal scaling factor
mostly depends on the number of integrations inside the correlator.

\begin{figure}
  \centering
  \includegraphics[width=100mm]{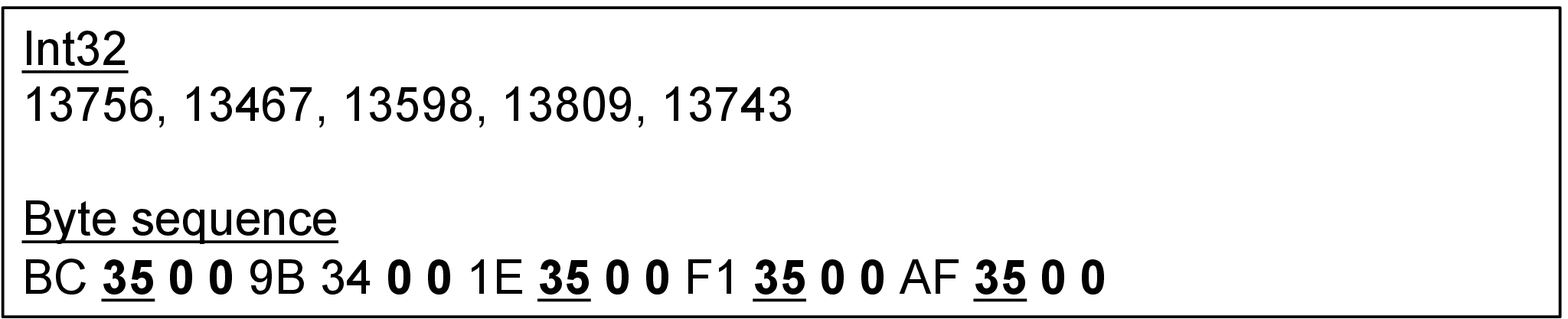}
  \caption{Rounded and converted to int32 fragment of output of MWA correlator. The second line represents 
  the bytes how these numbers are stored in a computer. The bytes designated in bold or underlined can be
  efficiently compressed with a binary algorithm.}
  \label{fig:int}
\end{figure}

Then the values are rounded as per "half round up" method and converted to int32 data type. 
As it can be seen from Figure~\ref{fig:int}, there are now many more bytes that a binary compression is able to compress.

\emph{cfitsio} \citep{white2012tiled} library had been used to compress the resampled data due to the fact that 
the correlator outputted the data as a set of FITS \citep{1979ipia.coll..445W} files. Equally, other binary compression software 
could be used to achieve the compression, but due to the fact that \emph{cfitsio} was already used, the library has allowed 
the compression and decompression
to be done completely transparently for the end user -- no other software had to be changed. 

\emph{RICE} and \emph{GZIP2} compression options available in \emph{cfitsio} had been tested with \emph{RICE} been found 
performing slightly better for optimal scaling factors. Table~\ref{tab:ratios} shows the typical compression ratios, as well as 
the percentage of affected data in tests. The "affected" is referring to such a data which uncompressed value is not equal to the original value.

\begin{table}[ht]
\centering
\begin{tabular} {|l|l|l|}
\hline
Scaling factor & Data affected (\%) & Compression Ratio (RICE) \\
\hline
1 & 1.35 & 1:3.1\\
4 & 1.52 & 1:2.6 \\
10 & 1.52 & 1:2.3 \\
100 & 1.08 & 1:1.9 \\
1000 & 0.23 & 1:1.6 \\
\hline
\end{tabular}
\caption{Typical compression ratios}
\label{tab:ratios}
\end{table}

\section{Conclusion}
\emph{uvcompress}  algorithm achieves high compression ratio what has allowed MWA 
to reduce the storage size. 

\bibliography{P1-8}

\end{document}